\RequirePackage{fix-cm}

\documentclass[twocolumn]{svjour3}          
\usepackage{float}
\smartqed  
\usepackage[singlelinecheck=false]{subcaption}

\usepackage[colorlinks=true, allcolors=blue]{hyperref}
       
\usepackage{graphicx,natbib,url}
\usepackage{aas_macros}

\begin{document}

\title{Development of a one-dimensional position sensitive detector for Compton X-ray polarimeters}

\titlerunning{NaI(Tl) scintillator characterisation}


\author{Abhay Kumar$^{1,2,*}$  \and
	Santosh V. Vadawale$^{1,}$ \and 
	N. P. S. Mithun$^{1}$ \and
	Tanmoy Chattopadhyay$^{3,}$ \and 
	S. K. Goyal$^{1}$ \and
	A. R. Patel$^{1,}$ \and 
	M. Shanmugam$^{1}$} 

\authorrunning{Kumar et al.} 

\institute{
$^1$  Physical Research Laboratory, Astronomy \& Astrophysics Division, Ahmedabad, India, 380009 \\
$^2$ INAF$-$Istituto di Astrofisica e Planetologia Spaziali di Roma, Via Fosso del Cavaliere 100, 00133 Roma, Italy \\
$^*$Corresponding author: Abhay Kumar\\
Tel.:+39 06 4993 4098\\
\email{abhay.kumar@inaf.it}\\ 
$^3$ Kavli Institute of Particle Astrophysics and Cosmology, Stanford University 452 Lomita Mall, Stanford, CA 94305, USA}

\date{Received: date / Accepted: date}

\maketitle

\begin{abstract}

The scientific potential of X-ray polarimetry has long been recognized, but the challenges in measuring polarization have left it largely unexplored, particularly in the hard X-ray regime. While tremendous advancement has been made in soft X-ray polarimetery, the lack of sensitive hard X-ray polarimeters and polarisation measurements continues to limit our understanding of high-energy astrophysical processes. With the development of hard X-ray mirrors, it is now possible to develop a sensitive focal plane hard X-ray polarimeter. One such effort is Compton X-ray Polarimeter (CXPOL), a prototype developed at Physical Research laboratory, India, which consists of a plastic scintillator as active scatterer readout by Photomultiplier Tube (PMT) surrounded by CsI(Tl) scintillators in cylindrical array with Si photomultiplier (SiPM) readout from one side. First results of the prototype have been demonstrated in 20 to 80 keV energy range \citep{chattopadhyay15}. The sensitivity of the instrument can be significantly enhanced using faster and better light yield scintillator like NaI as absorbers. 
Further, the use of a position-sensitive scatterer surrounded by position-sensitive absorbers, can also provide spectroscopic information by measuring the interaction position along the length and from the known energy depositions in the detectors. Position sensitive detectors are also helpful in mitigating the systematic effects introduced by the off-axis events in the polarisation measurements. Here, we demonstrate the detection sensitivity in the 100x20x5 mm$^{3}$ NaI(Tl) scintillator absorber readout on both ends by Silicon Photomultiplier (SiPM) arrays operating in co-incidence. In this work, we characterize the first prototype of this detector system and investigate the variation in energy and position resolution, and light output with irradiation position along the length of the detector. The two end readout in co-incidence also reduces the overall SiPM background per absorber by an order of magnitude, further enhancing the polarimetric sensitivity of the instrument.


\end{abstract}

\section{Introduction} \label{sec-Intro}

Since the birth of X-ray astronomy, the importance of polarimetry has been well recognized, but sensitive polarization measurements were lacking until recently due to its inherent measurement challenges. Measurement of polarization provides two additional parameters $\--$ polarisation degree and angle, which are related to the level of asymmetry and orientation of the source, respectively.

In the early days of X-ray astronomy, there were a few attempts to measure X-ray polarization at low energies ($<$ 10 keV) in the 1970s with a few space-borne and balloon-borne experiments\citep{weisskopf76,weisskopf78,angel69,wolfe70,novick72,griffiths76,gowen77,silver79,hughes84}; but these yielded upper limits at best, except for Crab and Cygnus X-1. 
After these initial efforts, no dedicated experiments were carried out for over three decades, until the Imaging X-ray polarimeter (IXPE) \citep{weisskopf2021} which was launched in 2021, and advanced the field of X-ray polarimetry by providing sensitive polarisation measurements in 2-8 keV which enabled the deeper understanding of the astrophysical sources. IXPE provided first of its kind polarisation measurements of several classes of X-ray sources, both galactic \citep{krawczynski2022,Taverna2022,Xie2022} and extragalactic \citep{Liodakis2022,Marin2023}. However, these measurements are limited to the soft X-ray regime. Energy and state dependent polarisation measurements over a broader energy band, enabled through simultaneous spectro-polarimetry, can reveal the contribution of different emission components and provide further insights on the geometry and emission mechanism of the radiation \citep{krawczynski2022,chattopadhyay2023_cygx1}. The number of measurements in the hard X-rays are only handful e.g. POGO+~\citep{chauvin18a,chauvin18,friis2018}, X-calibur~\citep{Abarr_2020,awaki2025}, INTEGRAL~\citep{dean08,forot08,jourdain12,laurent11}, AstroSat~\citep{vadawale17,chattopadhyay2022,chattopadhyay2023_cygx1} and most of these measurements are significantly limited by poor statistics. This is primarily due to low sensitivity of the instrument, further aggravated by the scarcity of photons in the hard X-rays.

The problem of scarcity of the photons in hard X-rays can be mitigated by the development of hard X-ray optics such as that used in the NuSTAR and Hitomi \citep{harrison13,takahashi2016}. With the availability of focussing optics in hard X-rays, it is also now possible to conceive compact focal plane detector geometries that are optimized for utmost polarimetric sensitivity and a significant reduction in the background due to narrow field of view of the telescopes \citep{chattopadhyay2021_review,bernard2024gamma}. 
The importance of hard X-ray optics in sensitive polarization measurements has been demonstrated recently by the XL-Calibur \citep{abarr2021} which is a balloon borne instrument with hard X-ray focusing optics. XL-Calibur measured polarization of Crab with an unprecedented 8.6$\sigma$ detection significance with 49.7 ks observation \citep{awaki2025}.

In this context, Physical Research Laboratory (PRL), India, has been working on the development of a compact focal plane Compton X-ray polarimeter (CXPOL) operating in the energy range of 20 to 80 keV to demonstrate and prepare readiness level for potential astronomical missions from India. The design and the first results of the prototype instrument have been reported in \cite{chattopadhyay13,chattopadhyay14_cxpol,chattopadhyay15}. This prototype instrument used a plastic scatterer and a cylindrical array of sixteen CsI(Tl) scintillator as absorber. The plastic scatterer was readout by a Photomultiplier tube (PMT) 
and each of the CsI(Tl) detectors \citep{hofstadter1950properties} was readout by a single silicon photomultiplier~\citep{buzhan2003silicon,yeom2013fast} at one end. The prototype demonstrated the intended proof-of-concept of the polarimetric configuration and the readout electronics. It also brought up a few aspects of improvement in the sensitivity as discussed in \citet{chattopadhyay15}. For example, it was noted that the CsI(Tl) detectors are only sensitive to few centimeters from the SIPM end, due to poor light collection. It was proposed that the use of a faster scintillator with dual-end readout by low-noise SiPMs will provide sensitivity throughout the length of the scintillator and simultaneous spectral information using Compton kinematics, provided interaction positions in the scatterer and absorber are known (Kumar, 2023, thesis). The similar concept of Compton polarimeter can also be extended for instruments with wide field of view with collimators, by making appropriate changes in the scatterer and absorber detector geometries (Mithun, 2024, thesis). It is pointed out that for the wide field instrument, the azimuthal response of the polarimeter is affected by the off-axis events above several degrees from the pointing direction \citep{yonetoku06, Muleri2014}.  The position sensitive absorbers and scatterers enables reconstruction of the scattering angle which helps in reducing systematics introduced by off-axis events from a known source position \citep{Muleri2014}. This approach also enhance the polarimetric sensitivity by weighting events which are scattered close to 90 degrees (events having larger response to polarisation).

In this paper, we present the next version of scintillator modules as absorbing detectors of the Compton polarimeter. The detector module consists of a NaI(Tl) scintillator read out by SiPMs on both ends. In addition to providing X-ray detection sensitivity over the entire length of the scintillator, two end readout of the scintillator also provides interaction position measurement of the incident photons using  the ratio of signals on the opposite ends. In the following sections, we present the motivation and detector design for the revised absorber configurations. Section~\ref{sec:electronics_expsetup} and \ref{sec:results} describe the test set up and the results respectively. In sec.~\ref{sec:summary} , we summarize the work and discuss the future plans.

\section{Revised absorber configuration: Motivation and Design}\label{sec:motivation}

\subsection{Motivation}

The CXPOL prototype~\citep{chattopadhyay15}
used 15 cm long CsI (Tl) scintillator bars with one ended SiPM readout as absorbers. The detectors were found to be sensitive only close to the SiPMs. The combination of a long scintillator decay time and one-end readout also led to a non-uniform gain along the detector. We explored the possibilities to improve the sensitivity of the detector with the use of a new generation scintillator and SiPMs as well as to also obtain additional information useful for spectroscopy.  

The simultaneous spectroscopic information of the incident radiation is equally important as the measurement of polarization. The energy of the incident photon can be measured in two ways:\\
(a) By the sum of the energies deposited in the scatterer and the surrounding absorbers. However, spectral information is severely limited by the poor energy resolution in plastic scintillators in this case.\\ 
(b) Another way is by using Compton kinematics, where the source photon energy is measured from the polar scattering angle and the energy deposited in the absorber.  
For the measurement of scattering angle, the position of interaction in the absorber and the scatterer is required. The scattering angle can be estimated by a simple geometry, knowing the interaction position of the photon along the length (1D) in absorber and scatterer. 

To determine the interaction position within the absorber detectors, one approach is to employ two-end readout, where the position of interaction can be measured from the ratio of signals generated in the SiPMs by the scintillation photons at the two ends of the scintillator. This configuration also enhances sensitivity across the entire length of the scintillator, overcoming the limitations of the earlier detector design. Based on these considerations, the detailed absorber detector design and the selection of scintillators and SiPMs are discussed in the next section.

\subsection{Detector Design}\label{sec_detect_design}

In the revised design, CsI(Tl) scintillator has been replaced with NaI(Tl), which offers a fourfold improvement in light decay time constant while maintaining comparable light yield (see Table \ref{table_scint_parm}). The CsI(Tl) detectors were read out using KETEK SiPMs (PM3350-EB) having a background level of $\leq$ 500 kHz/mm$^{2}$ , which is significantly higher than that of the new generation SiPMs. We replaced the KETEK SiPMs with SensL (now OnSemi) MicroJ-series (60035) SiPMs having lower background (factor of five lower), improved photo-detection efficiency, better recovery time of the micro-pixels, and a spectral response well-matched with the peak emission of NaI(Tl) (see Table \ref{table_simp_parm}).

\begin{table*}[!ht]
	\begin{center}
		
		\caption{Characteristics of NaI(Tl) and CsI(Tl) scintillator for the X-ray detection.}
		
		\label{table_scint_parm}
		
		\begin{tabular}{ l c c  }
			\hline
			Parameter & NaI(Tl) [1][2]  & CsI(Tl) [1][2]  \\
			\hline
			Decay constant (ns) & 230 & 1050  \\
			Density (g/cm$^{3}$)& 3.67& 4.53  \\
			Radiation length (cm) & 2.59 & 1.86 
			\\
			Emission peak (nm) & 415 & 540  \\
			Light yield (Ph/MeV) & 40,000 & 48,000  \\
			
			\hline
		\end{tabular}

		\begin{flushleft}
			$[1]$ More details are given in \citet{shionoya2018,tamagawa2015,koppert2019comparative}\\
			$[2]$ \url{https://luxiumsolutions.com/sites/default/files/2021-09/Scintillation-Materials-and-Assemblies.pdf} and \url{https://www.advatech-uk.co.uk}\\
			
		\end{flushleft}
	\end{center}
\end{table*}

\begin{table*}[!ht]
	\begin{center}
		
		\caption{Some important characteristics of the SensL J-series and KETEK  SiPM at V$_{br}+\sim2.5$V and 21${^\circ}$C}
		
		\label{table_simp_parm}
		
		\begin{tabular}{l | c c |c  }
			\hline
			{Parameters} & \multicolumn{2}{c|}{SensL (now OnSemi) [1](MicroJ-series)}  & {KETEK [2](PM3350-EB)}  \\
			\cline{2-3}
			
			& 6 mm  & 3 mm &  \\
			
			& (60035) & (30035) &  \\
			
			\hline
			
			Active area ($mm^{2}$) & $6.07\times6.07$& $3.07\times3.07$ & $3\times3$ \\
			
			$V_{br}$ (V)&{24.7(max)} &{24.7(max)}&25 \\

			Microcell size ($\mu m$) &35&35 & 50\\
			No. of microcells &22,292&5676&3472\\
			Peak PDE ($\%$) &38&38 & 28\\
			Peak wavelength (nm)& 420& 420 & 430 \\
			Dark count rate (kHz/$mm^{2}$) &50&50& 250 \\
			Gain ($\times 10^{6}$)&2.9&2.9&3.6\\
			Recovery (ns)&50&45&130\\
			
			\hline
		\end{tabular}

		\begin{flushleft}
			$[1]$ \url{https://www.onsemi.com/pdf/datasheet/microj-series-d.pdf}\\
			$[2]$ \url{https://www.ketek.net/wp-content/uploads/2017/01/KETEK-PM3325-EB-PM3350-EB-Datasheet.pdf}\\

		\end{flushleft}
	\end{center}
\end{table*}

\subsubsection{Optimization of detector shape and size}

\begin{figure*}[ht!]
	\centering
    \includegraphics[scale=0.6,trim={0.0cm 0.cm 0.cm 0.0cm},clip]{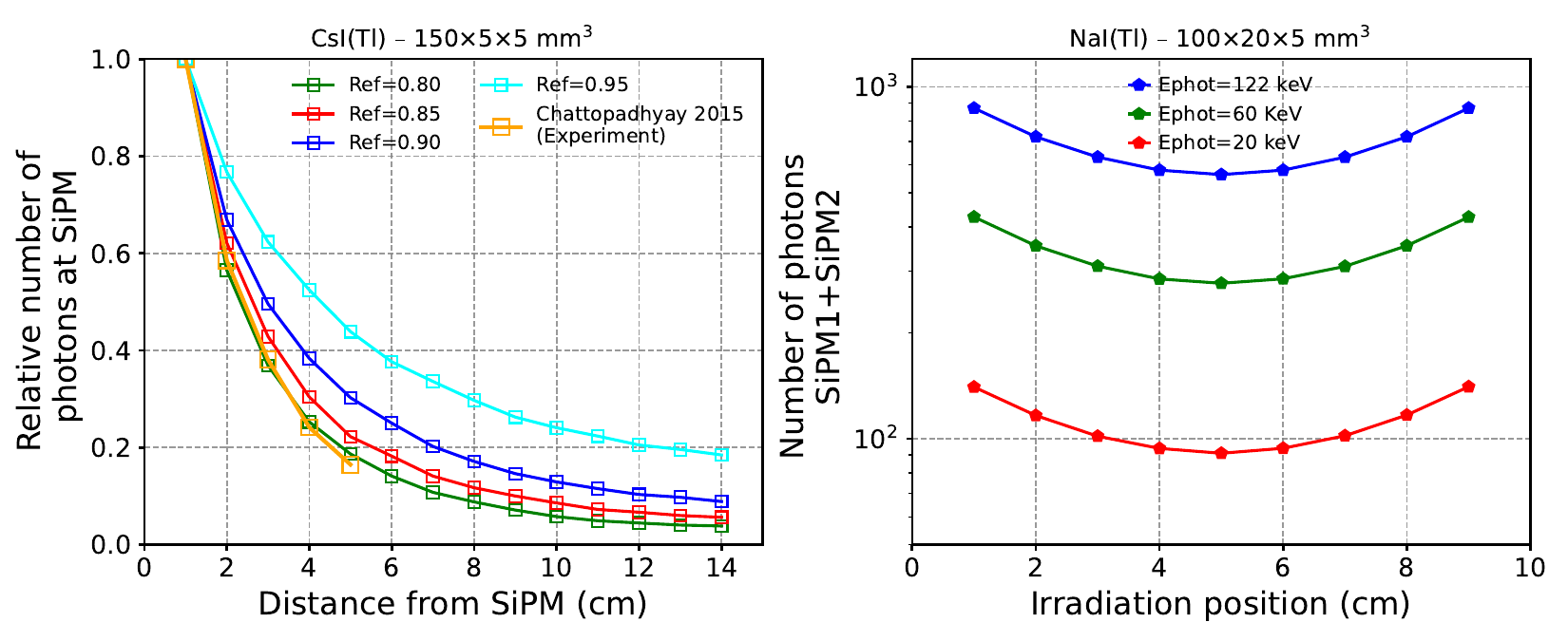}
		
	\caption{Left figure shows the variation of relative number of photons at SiPM with distance considering different reflectivity in simulations and overplotted with the experimental values for CsI(Tl) scintillator absorber of size (150$\times$5$\times$5 mm$^{3}$) with one end readout by a SiPM. Right figure shows the variation of the number of photons with irradiation position at 20, 60 and 122 keV for NaI(Tl) scintillator absorber of size (100$\times$20$\times$5 mm$^{3}$) with two end readout by the SiPMs.}%
	\label{fig:reflectivity}
\end{figure*}

Besides using faster scintillators and two-end SiPM readout, it is equally important to optimize the length, width and thickness of the NaI(Tl) scintillators to maximize the light collection efficiency. The current detectors are 2 cm in width and 5 mm thick. The increased width (therefore enhanced cross-sectional area) ensures better light collection compared to the narrower detectors used in previous version. The 5 mm thickness, consistent with the previous detector version, allows proper coupling to the SiPM array and offers high absorption efficiency for hard X-rays. The length of the scintillator is the most critical parameter to optimize. While a longer detector geometry will ensure higher polar scattering angle coverage for the scattered events, the detector must also maintain sensitivity along its length and achieve the lowest possible energy threshold in the desired energy range of operation. We decided the length based on insights gained from the previous CXPOL test results and further optimized through detailed Geant4 simulations. Because, CsI(Tl) absrobers were sensitive up to 4-5 cm from the readout end, a dual-end readout of a 10 cm long scintillator is expected to provide detection sensitivity throughout its length. To verify this, we carried out detailed optical Monte carlo simulations in Geant4 for a NaI(Tl) scintillator. One of the key parameters for such simulations is the reflectivity of the scintillator wrapping, typically made of polytetrafluoroethylene (PTFE). The intrinsic reflectance of PTFE is $>$95$\%$ across visible range \citep{barnes1998}. However, the effective reflectance in a packaged detector is often lower due to wrapping non-idealities such as surface roughness and imperfect optical contact. For the simulations conducted before NaI(Tl) procurement, we used an effective reflectivity of 80$\%$  as a realistic prior. The reflectivity is determined from simulations of a CsI(Tl) detector (same as the CXPOL v1), where simulation were done for a range of reflectivities and compared with the experimental measurements. The best agreement is achieved for a reflectivity of 80$\%$ (see Figure \ref{fig:reflectivity}).

In the simulations, we define a NaI(Tl) scintillator (100$\times$20$\times$5 mm${^3}$) and its housing with 0.5 mm thick aluminum on three sides and a carbon window (0.2 mm thick) at the front to allow X-rays to enter while preventing optical photon leakage. The optical properties of the scintillator are defined as given in Table \ref{table_scint_parm}. The typical emission spectra of the NaI(Tl) scintillator is taken from the Advatech-UK website \footnote{\url{https://advatech-uk.co.uk/nai_tl.html}} and the corresponding refractive index at different wavelengths is calculated using the formula given by \citet{li1976refractive}.
Optical simulations are carried out at 20, 60, and 122 keV. The scintillator is irradiated at nine positions spaced 1 cm apart along its length with one million X-ray photons at each point. The number of optical photons reaching the SiPMs are recorded on both ends of the scintillator for each positions.
Based on these simulation results (see Fig.~\ref{fig:reflectivity}), the NaI(Tl) detector is expected to remain sensitive along its entire length for incident energies starting from 20 keV, provided that the equivalent background level is kept below $\sim$ 100 optical photons.

\subsubsection{Assembled detector module}

Due to the hygroscopic nature of NaI(Tl), we procured an enclosed detector module from Advantech, UK, which includes a NaI(Tl) scintillator bar optically coupled with SiPMs at both ends. The crystal is wrapped in polytetrafluoroethylene (PTFE; a.k.a Teflon)  on all faces except for the two ends. The pre-wrapped crystal is housed in an aluminium enclosure with a minimal air gap to allow smooth insertion of the crystal into the housing. The aluminum case encloses three sides of the scintillator, while the two ends are open for scintillation photon readout. The front face of the case is covered with a 0.2 mm thick carbon window.

\begin{figure*}[ht!]
	\centering
	\begin{subfigure}{.35\linewidth}
		\includegraphics[scale=0.5,trim={0.1cm 0.1cm 0.cm 0.cm},clip]{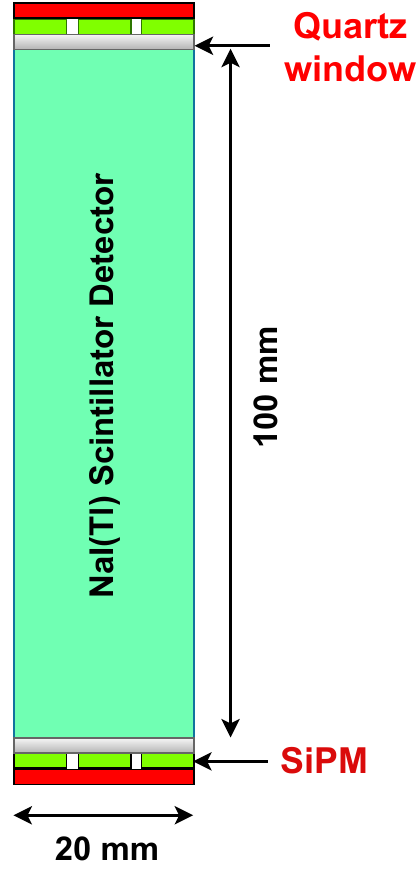}
		\caption{(a)}
	\end{subfigure}
	\begin{subfigure}{.5\linewidth}
		\includegraphics[scale=0.85,trim={0.3cm 0.cm 0.cm 0.3cm},clip]{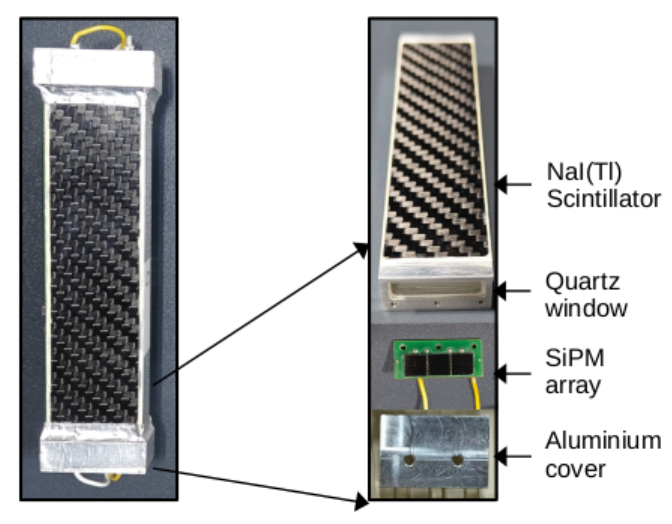}
		\caption{(b)}
	\end{subfigure}%
	
	\caption{The figure on the left is the schematic of the scattering geometry of the proposed spectro-polarimeter. The right side figure shows the schematic of the absorber detector (100$\times$20$\times$5 mm$^{3}$) with two end readout by the SiPMs.}%
	
	\label{fig:scint_detector}
\end{figure*}
Each end of the scintillator is coupled with an array of three SiPMs, each with an active area of 6.07$\times$6.07 mm$^{2}$ (see Fig.~\ref{fig:scint_detector}(b)). The size of the SiPMs is selected to have maximum collecting area to get better sensitivity of the detector. The SiPMs in each array are connected in series with an integrated current signal at the output. The SiPM array is coupled with the scintillator through a quartz window using a screw mounted assembly (see Fig. \ref{fig:scint_detector}). The quartz window is the part of the hermetic sealing to protect the NaI(Tl) from moisture. To minimize reflection losses, each optical interface (scintillator-quartz and quartz-SiPM window) is coupled using a 0.5 mm thick layer of optically clear silicone-based gel with a refractive index intermediate between the two adjoining materials.
To minimize stray light background, both ends of the scintillator - hereafter referred to as SiPM1 and SiPM2 are sealed with aluminum caps (see Fig.~\ref{fig:scint_detector}).

\section{Readout electronics and experiment set up}\label{sec:electronics_expsetup}

\begin{figure*}[ht]
	\centering
    \includegraphics[scale=0.25,trim={2cm 4cm 1cm 2cm},clip]{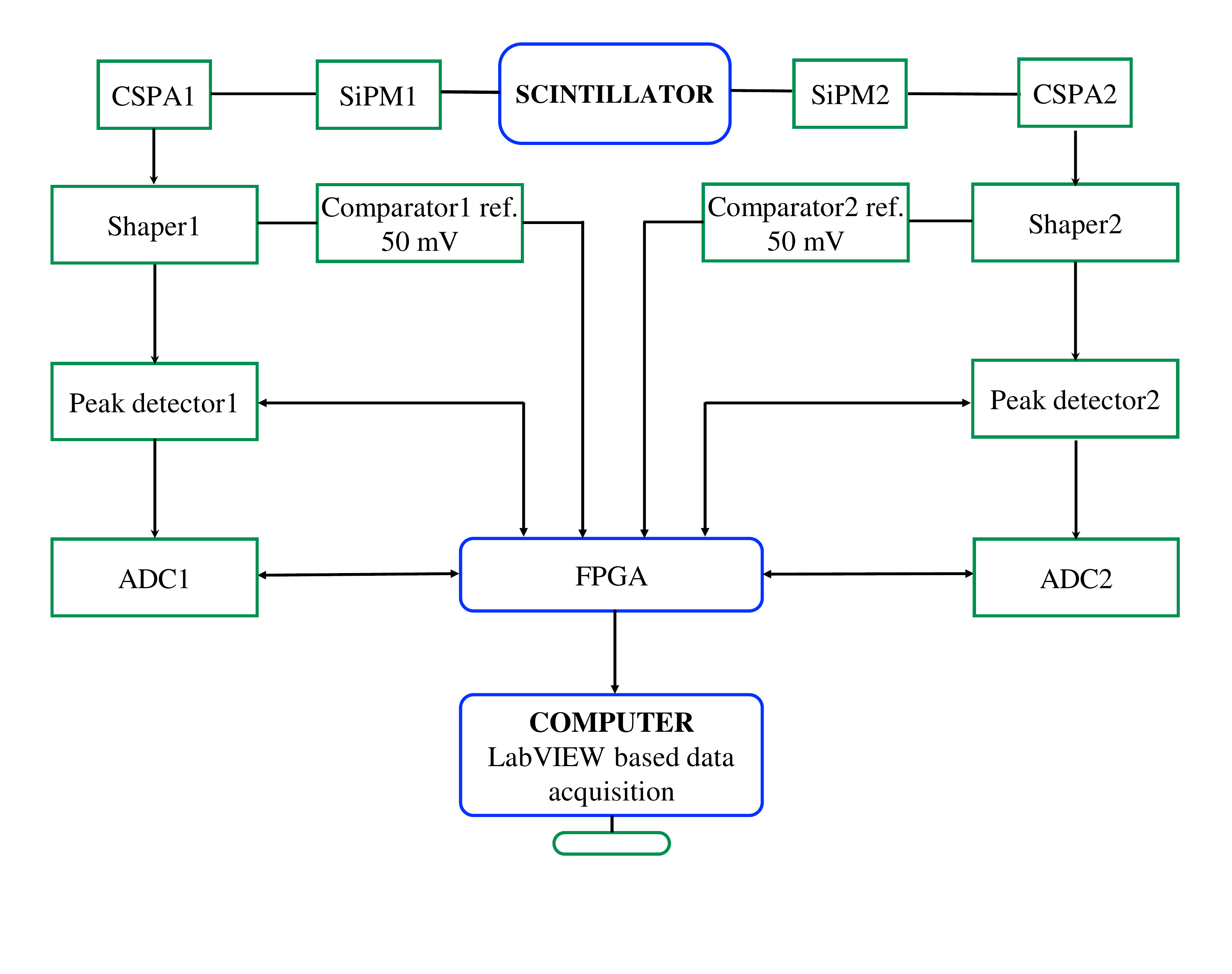}
	\caption{Block diagram of the electronic chain employed for the experimental setup}
	\label{1D_elect}
\end{figure*}

\begin{figure*}[ht]
	\centering

    \includegraphics[scale=0.55,trim={0.1cm 0.1cm 0.1cm 0.1cm},clip]{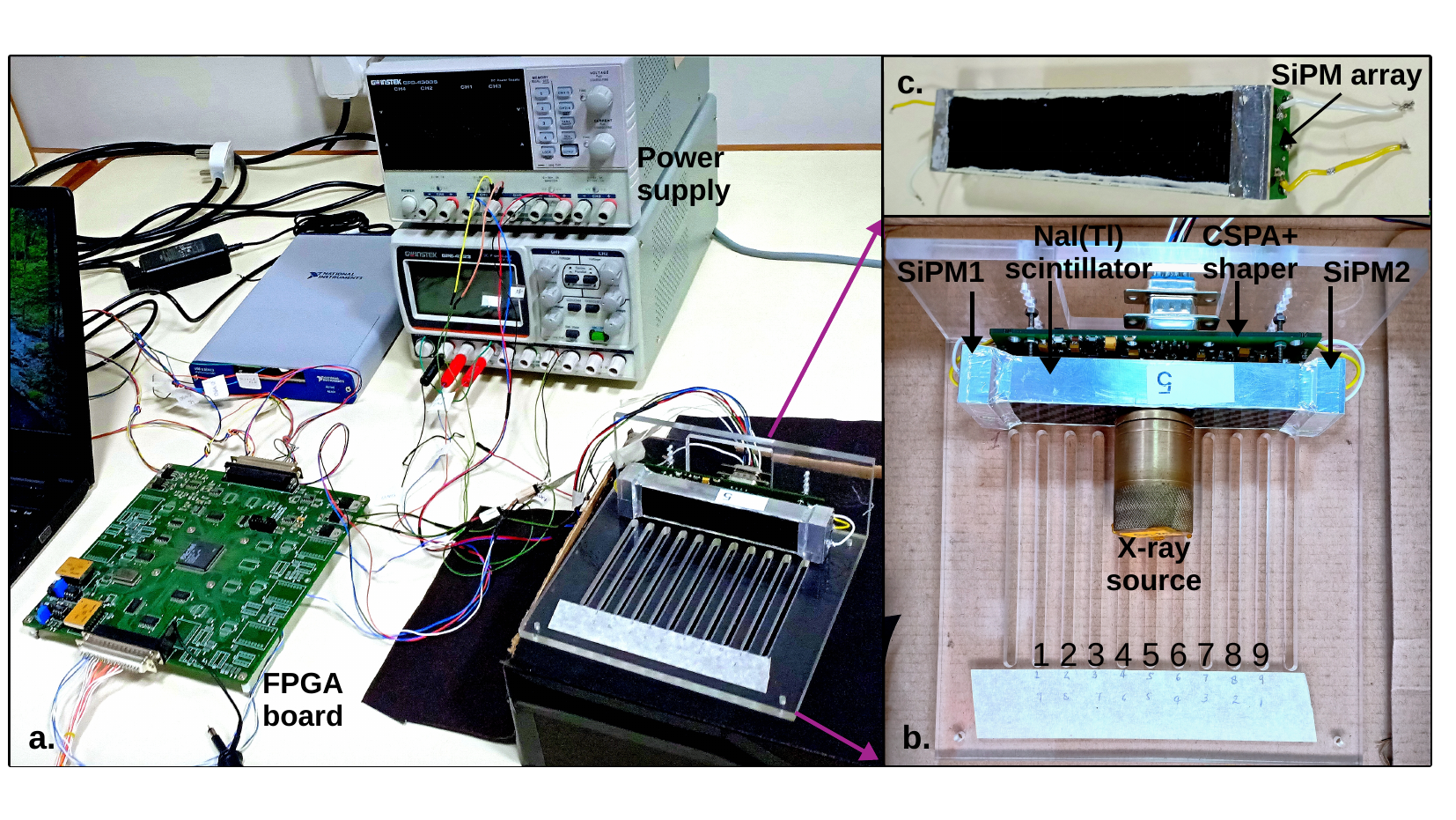}
	\caption{Panel a.: Experimental setup for the position measurement consisting of detector setup, front-end, back-end readout electronics, and power supply. 
		Panel b.: NaI(Tl) scinitllator is packaged in the aluminium cover with SiPM array at both ends, and CSPA+Shaper is shown. Panel c.: The bare scintillator readout by an array of 3 SiPMs at both ends is shown.}
	\label{expset}
\end{figure*}

We developed a custom readout electronics for the detector system, comprising two boards (see Fig.~\ref{expset}). The first board houses the charge sensitive preamplifiers (CSPAs) and shapers, which amplify and shape the analog signals from the SiPMs. The shaped signal from the two ends of the scintillator is fed to the second board, referred to as the FPGA board, which integrates a field-programmable gate array (FPGA), peak detectors, and  comparators for further signal processing and triggering. The processed signals from the FPGA board are then interfaced with a Data Acquisition System (DAQ) for recording. The block diagram is shown in Fig.~\ref{1D_elect}.

The output analog signals from the SiPMs are AC-coupled to an RC-feedback CSPA implemented with an LM6172IM operational amplifier (op-amp). CSPA integrates the input charge and produces a voltage pulse proportional to the number of optical photons detected by the SiPM. The CSPA output is then fed to a CR-RC-RC shaper that generates a Gaussian-shaped pulse with a peaking time of approximately 4 $\mu$s. When the shaped signal exceeds 50 mV, the comparator triggers the FPGA, which routes the shaper output to the peak detector. The 50 mV threshold was chosen to suppress triggers from electronic noise while maintaining sensitivity to genuine scintillation events, allowing reliable detection of both the 30 keV escape peak and the 60 keV photopeak in the two-end coincidence readout. The peak detector (AMPTEK PH300) holds the peak amplitude for 3.2  $\mu$s to allow analog-to-digital conversion (ADC) using a 12-bit ADC (ADC128S102). The peak detector then discharges in 1$\mu$s. Additional data processing which includes FPGA time stamping and data transfer takes 5 $\mu$s, resulting in a total electronics dead time of 13.2 $\mu$s per event. The peak detector, ADC, and shaper are discussed in more detail in \citet{patel2023}. The measured dead time of 13.2 $\mu$s corresponds to a maximum count rate handling capability of $\sim$ 76 kHz, which significantly higher than the expected count rates from even the brightest on axis sources in 20-100 keV (considering reasonable effective area). This ensures minimal dead time losses and low pileup probability.  
The events are recorded with 1 $\mu$s coincidence time window. When a trigger is generated from one SiPM, and a second signal from the other SiPM arrives within a 1 $\mu$s (discharge time window of the peak detector), both signals are readout. In absence of one of the SiPM outputs, only detected signal is readout with the other recorded as zero digital charge. The SiPMs are operated with a reverse bias of 26.5 V, approximately 2.5 V above their breakdown voltage of 24.2 V. For data acquisition, we developed a LabVIEW-based interface that records time-tagged events. Each event's ADC value and timestamp (with 1 $\mu$s resolution from the FPGA internal clock) are saved for further analysis. 

We carried out the experiments to investigate the detector performance at different interaction positions along the length of the scintillator. The experimental setup is assembled inside a light-tight dark box, and placed inside a darkroom to minimize the background contamination from stray light. The setup includes the NaI(Tl) scintillator detector, an Am$^{241}$ X-ray source (59.54 keV), and the readout electronics (see Fig.~\ref{expset}(b)). All components are mounted on a transparent perspex platform having equidistant vertical grooves, designed to precisely position the X-ray source in front of the scintillator. The Am$^{241}$ radioactive source in its enclosure having small collimator ensuring small opening angle is placed close to the scintillator to minimize divergence and illuminate only a small region ($<$1 mm spatially) of the scintillator at a time. The scintillator is scanned along its length at nine equidistant irradiation positions (see Fig.~\ref{expset}(b)). Due to the aluminum end caps covering both ends of the scintillator, the X-ray source could not be placed close to scintillator near both the ends. As a result, the first irradiation point is positioned 1.5 cm away from the end. The data is acquired at each of the nine positions, until a sufficient number of events are recorded to ensure good statistics (typically corresponding to about five minutes of acquisition per position). For each detected photon, we estimate the photon interaction position and reconstruct the energy of the photon. The coincidence logic is applied during the data analysis. A valid coincidence is defined when both SiPMs register ADC values greater than zero within 1 $\mu$s time window. To characterize the background level, we also acquired the data without radiating the scintillator with an X-ray source. The results of the measurements are discussed in the subsequent sections. 

\section{Results}\label{sec:results}

\subsection{SiPM gain correction}

\begin{figure*}[ht!]
	\centering
	\includegraphics[scale=0.55]{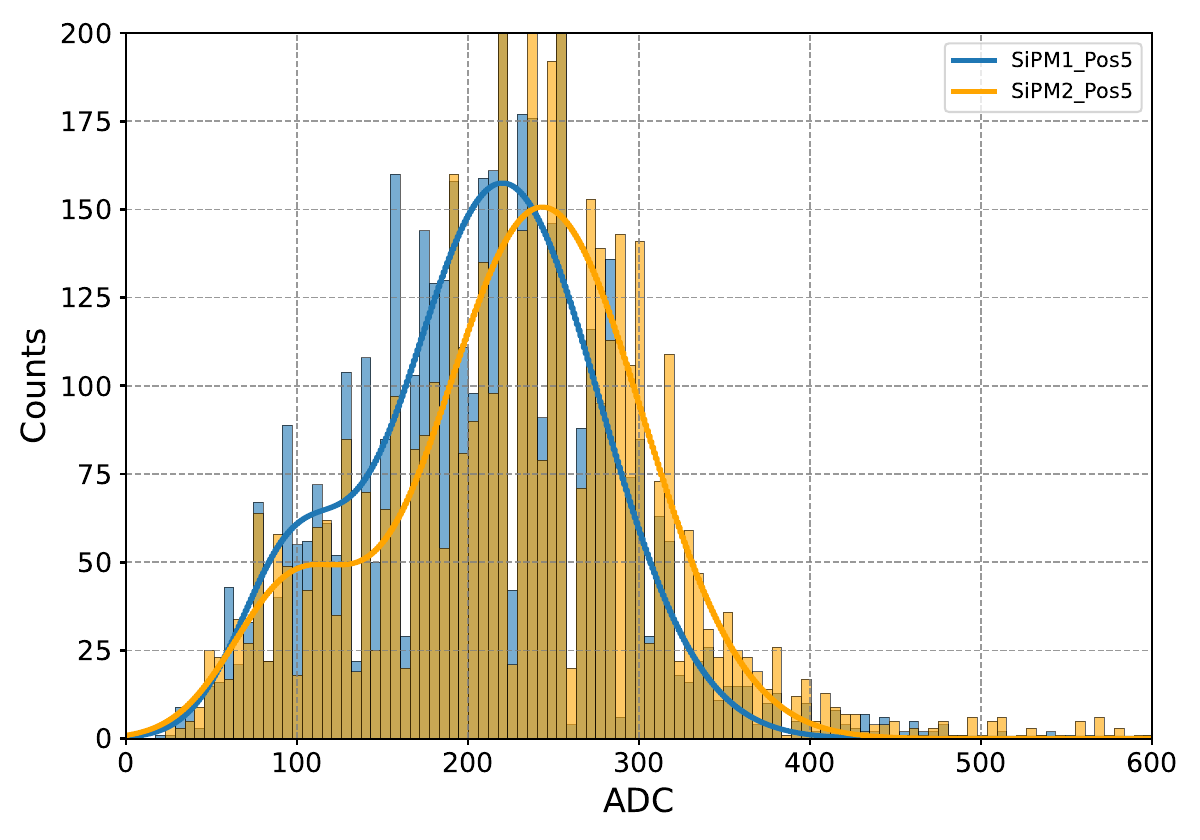}
	
	\caption{ADC distribution detected at two ends when irradiated at the center using an Am$^{241}$ X-ray source.}
	\label{fig:gainvariation}
\end{figure*}

We examined SiPMs output to assess any gain variation between two ends of the detector. Since we procured the scintillator detector as an enclosed package to ensure hermeticity, direct access to the SiPMs for calibration with a known source is not possible without compromising the optical coupling between the the quartz optical window and the SiPMs. To address this, we employed a symmetry-based calibration method: by irradiated the scintillator at its center with an Am$^{241}$ source and measured the coincident light output from both ends. Given that the optical path lengths from the center to both ends of the scintillator are equal, and assuming symmetric reflectivity on both sides, we expect similar signal amplitudes (ADC values) from both the SiPMs provided their gains are equal. The resulting signal distributions from the two ends are fitted using two Gaussians. SiPM2 signal distribution showed a higher peak value in the ADC  (hereafter ADC2) than the SiPM1 (hereafter ADC1), suggesting a gain difference (see Fig.~\ref{fig:gainvariation}). From the fitted peak positions, the ADC ratio (ADC2/ADC1) is found to be  $\sim$1.11$\pm0.03$. This gain variation is corrected in the subsequent analysis to ensure accurate energy and position measurements.

\subsection{Energy reconstruction}
The energy of the incident X-ray photon can be measured by the summing over the output signals obtained from the SiPMs on opposite ends. At each of the nine irradiation positions along the scintillator axis, we calculate the distribution of the ADC values (ADC1+ADC2) for the recorded events. These distributions are fitted with a Gaussian model to measure the centroid and the spectral width of the X-ray lines (Full Width at half Maximum / FWHM). 

\begin{figure*}[ht!]
	\centering
	
		\includegraphics[scale=0.55,trim={0.1cm 0.1cm 0.1cm 0.1cm},clip]{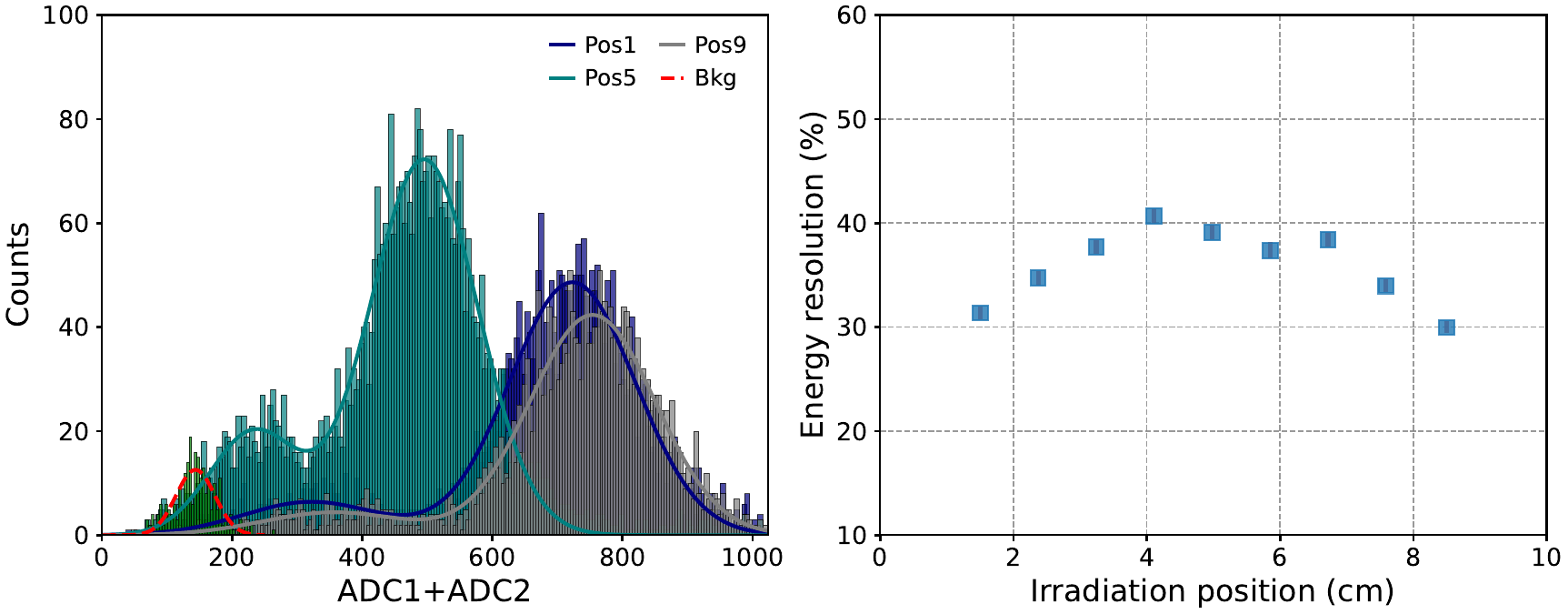}
	
	\caption{Left panel shows the Gaussian curves fitted to the profiles of ADC1+ADC2 at nine irradiation position is shown, and the energy resolution as a function of irradiation position is shown in the right panel for two-sided readout of the scintillator with SiPM. The Gaussian shown with red dotted line represents the background spectrum.}%
	\label{fig:ene_2end}
\end{figure*}

In coincidence mode of readout, the SiPM background events are significantly suppressed.  For each position, the distribution is fitted with a function consisting of two Gaussians where the first (lower-ADC) peak in the energy spectrum corresponds to the blend of iodine escape peaks (at 31.4 keV, 31.7 keV, and 27.7 keV), and the second peak represents the main 59.54 keV photopeak from the Am$^{241}$ source (see Fig.~\ref{fig:ene_2end}). Both the peaks are detectable along the entire length of the scintillator, suggesting the low energy threshold of this set up is at least $\sim$ 30 keV. We have also measured the NaI(Tl) background spectrum, where events are recorded in coincidence (shown by the red dotted line in Fig.~\ref{fig:ene_2end})).

\begin{figure*}
	\centering
	\includegraphics[scale=0.65,trim={0.1cm 0.1cm 0.1cm 0.1cm},clip]{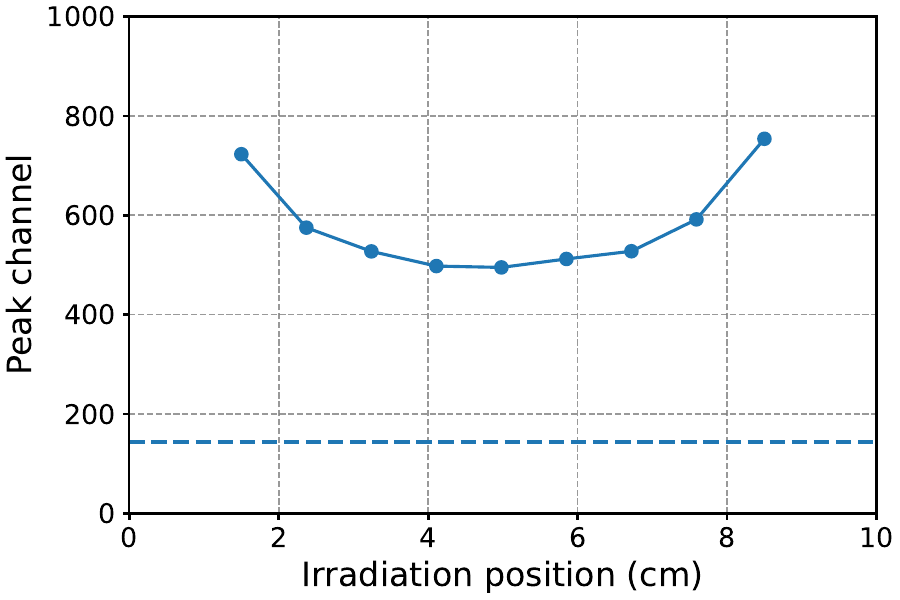}
	\caption{Peak channel as a function of irradiation position for two sided readout of the scintillator is shown.}%
	\label{fig:photo_peak}
\end{figure*}

The measured energy resolution at 59.54 keV is $\sim$ 30$\%$ on both ends and $\sim$ 40$\%$ at the center. When averaged over the entire length of the scintillator, we obtain a value of $\sim 34\%$ for energy resolution. The resolution degrades towards the center from either ends (see Fig.~\ref{fig:ene_2end}) due to reduced light collection at the center of the scintillator. Figure~\ref{fig:photo_peak} shows the variation of the 59.54 keV centroid as a function of the irradiation position, we see the centroid is highest close to the SiPMs and lowest at the center of the scintillator (see Fig.~\ref{fig:photo_peak}). This is consistent with the reduced light collection at the center due to increased optical path length leading to higher optical light difussion or scattering losses.

\begin{figure*}[ht!]
	\centering
	\includegraphics[scale=0.65]{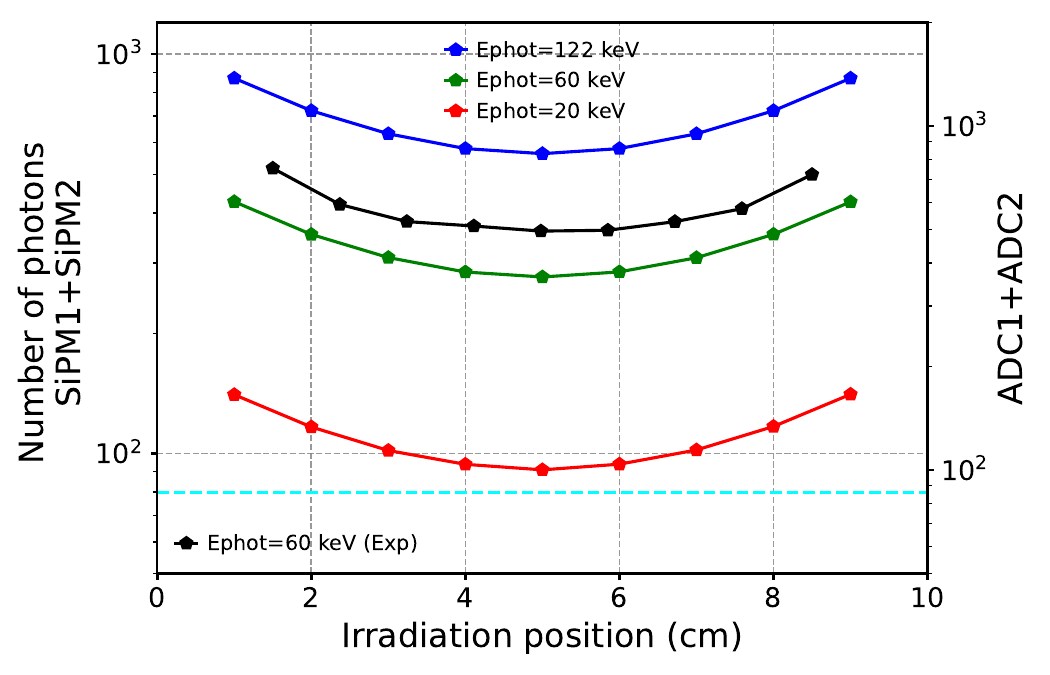}

	\caption{The variation of the number of optical photons detected with distance from the SiPM in tow end readout of NaI(Tl) scinti  llator at 59.54 keV are shown. The left side y-axis shows the simulation values, while the experimental values are shown on the right y-axis.}
	\label{exp_sim_comp}
\end{figure*}

We compare the experimental results with the Geant4 based optical simulation results discussed in Section~\ref{sec_detect_design} in Figure~\ref{exp_sim_comp}. The number of photons collected by the SiPMs for three different energies (122, 60, and 20 keV) are plotted against the left axis, whereas the experimentally measured 59.54 keV centroid are plotted against the right axis. We convert the background peak centroid (the red dotted line in Fig.~\ref{fig:ene_2end}) to equivalent number of optical photons (cyan horizontal line in Fig.~\ref{exp_sim_comp}) by multiplying it with the ratio of number of photons collected by SiPMs for 59.54 keV to the corresponding centroid in ADC.

The mean background level is seen to be around $\sim$ 100 optical photons (corresponding to the lower limit of the 20 keV curve), indicating that the detector retains sensitivity to energies above 20 keV. This can be further enhanced through several design optimizations. Optimizing the coincidence timing window and trigger threshold would suppress random coincidences and improve background rejection, while enhanced optical shielding at both ends of the scintillator would reduce photon leakage. Additionally, applying a high-reflectivity wrapping would increase internal light reflection and thereby improve light collection efficiency. Finally, replacing discrete front-end electronics with fast ASIC-based readout system with reduced parasitic and lower readout noise will allow further optimization in the integration time constants, enhancing the signal to noise ratio of the system.

\subsection{1-D localisation of individual X-rays}
\begin{figure*}[!ht]
	\centering
	
	\includegraphics[scale=0.6,trim={0.1cm 2.5cm 0.2cm 4cm},clip]{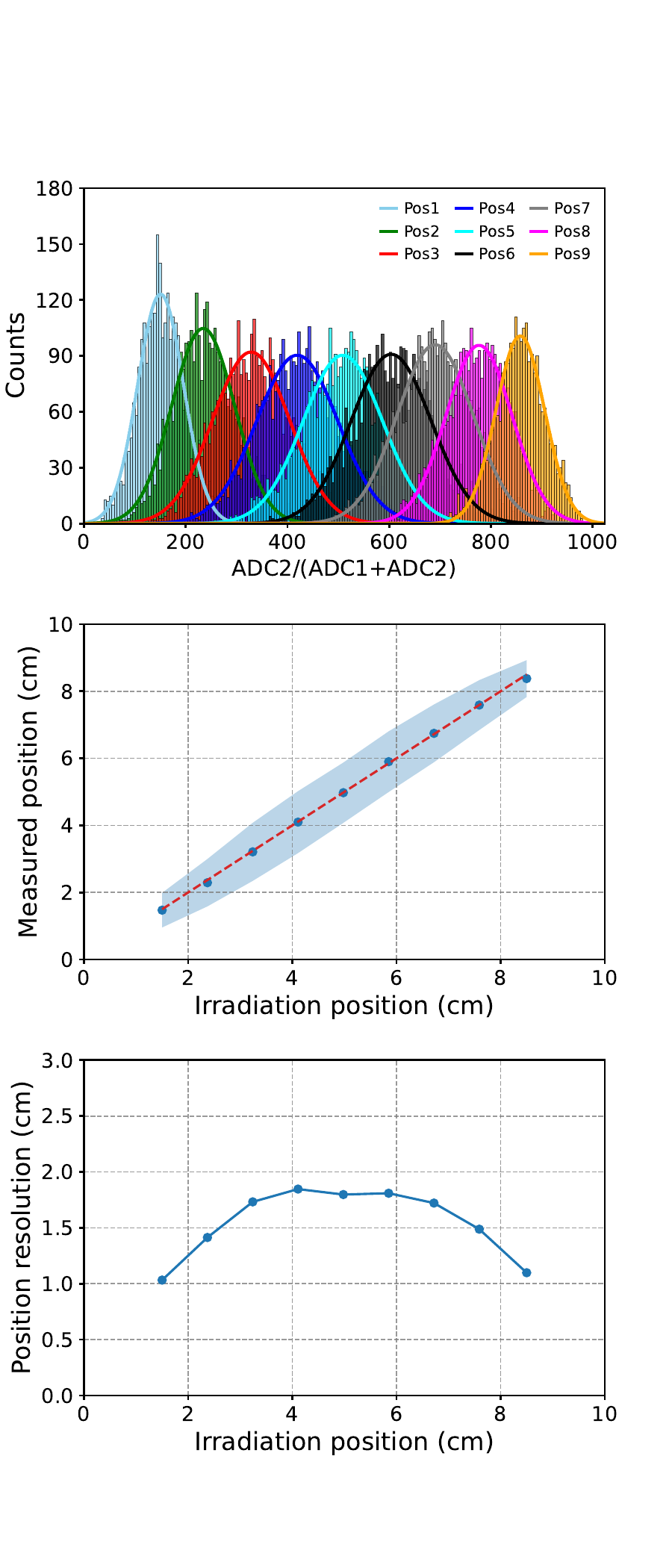}
	
	\caption{Top panel: Gaussian fits to the signal ratio profiles from the two-end SiPM readout of scintillator at nine irradiation positions is shown. Each color corresponds to a specific irradiation position, as indicated in the legend. Middle panel:red dashed line represents the expected trend between irradiation position and measured position (green markers).The shaded blue region represents the position resolution at each point.  Bottom panel: variation in resolution along the length of the scintillator is shown which is same as shaded blue region in middle panel.}
	\label{fig:position}
	
\end{figure*}

The interaction position of the X-ray photons along the scintillator axis (1-D localization) is determined by the ratio of the number of scintillation photons detected by SiPMs on either ends (ADC2/(ADC1+ADC2)). This ratio of the ADC values is scaled by a factor of 1024 for better visualization (see Fig. \ref{fig:position}) (top). For each irradiation point (position 1 to position 9), the distribution of the measured ratios for a large number of events is fitted with a Gaussian to extract the centroid location and the FWHM of the distribution. The centroid indicates the mean interaction position which is expected to coincide with the true irradiation position. The shaded region denotes the FWHM (Fig.~\ref{fig:position} middle) which quantifies the spatial resolution of the detector (Fig. \ref{fig:position} bottom). For this analysis, we selected events corresponding to the 59.54 keV centroid of Am$^{241}$. We exclude the iodine escape peaks from the analysis, which otherwise broadens the position distribution and degrades the spatial resolution.

We find the measured localization is well aligned with the irradiation location with an average positional resolution of 1.55 cm. Near the SiPMs, the position resolutions are measured to be 1.04 cm and 1.09 cm. The strong signal contrast between the opposite ends allows more precise estimation of the interaction position close to the SiPMs compared to center. It is to be noted that the measured positional resolution includes contribution of the finite beam size of the radioactive source. It is estimated to be $\sim$ 0.2 mm considering the known aperture of the source enclosure and the distance between the source and the scintillator. The measured position resolution, however, is dominated by the intrinsic position resolution.

\subsection{Detection Efficiency $\--$ axial variation of coincidence events}

\begin{figure*}[ht!]
    \centering
    {\includegraphics[width=0.6\textwidth,trim={0.1cm 0.1cm 0.1cm 0.1cm},clip]{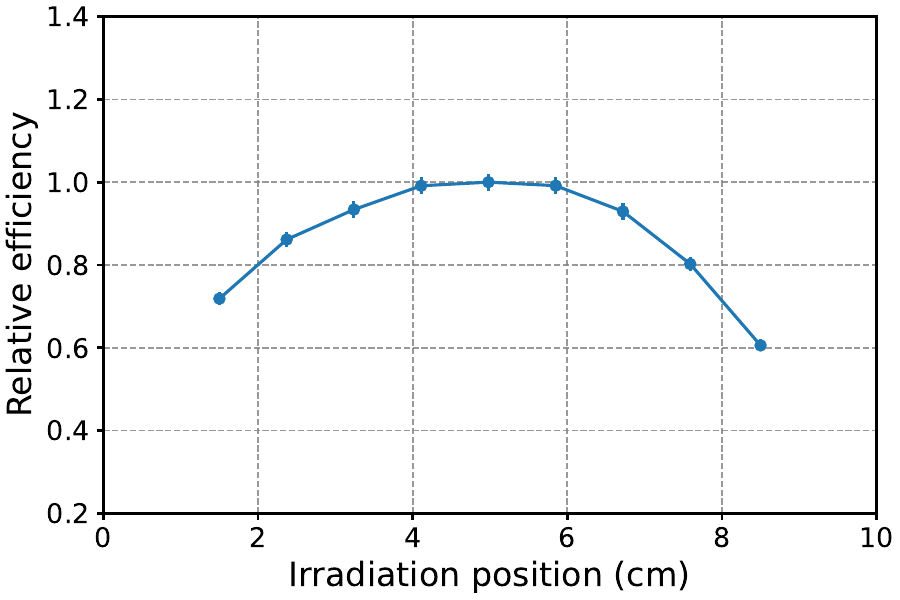} }
\caption{Relative efficiency as a function of irradiation position in two end readout coincidence mode}%
    \label{fig:eff}
\end{figure*}

To evaluate the detection efficiency of the NaI(Tl) scintillator with two-end SiPM readout, we measured the total number of valid coincidence events recorded at each irradiation position along the scintillator length. For each irradiation position, the relative efficiency is calculated as the ratio of the total valid coincidence events to the maximum number of valid coincidence events observed across all positions. Figure~\ref{fig:eff} shows that the relative efficiency is non uniform across the length of the scintillator. It peaks at the center and gradually declines towards both ends. The maximum efficiency at the center is observed due to the symmetric distribution of scintillation light. At this position, optical photons have approximately equal path lengths to both SiPMs, resulting in equal light sharing and a higher probability of exceeding the detection threshold at both ends, thereby high coincidence rate. The relative efficiency decreases as we move away from the center of the scintillator, dropping by approximately 28$\%$ on one end and up to 40$\%$ on other end with respect to the center. This asymmetry in efficiency at the two ends is attributed to the difference in optical coupling, reflectivity, and SiPM gains which affect light collection at both the ends.

\subsection{SiPM background}\label{sec_bkg}

\begin{figure*}[ht!]
	\centering
	\includegraphics[scale=0.8]{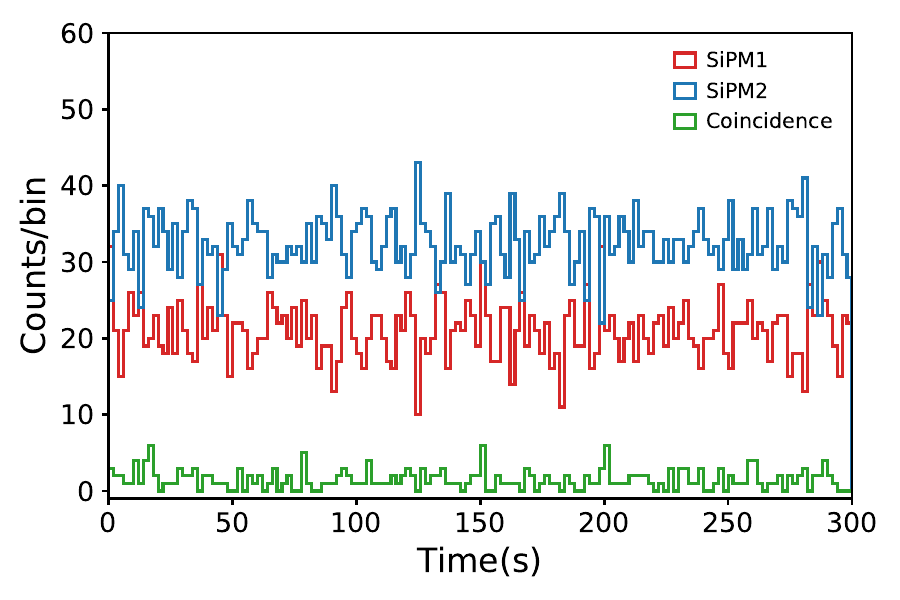}

	\caption{Background counts per (2-second) bin: SiPM1 is shown in red, SiPM2 (opposite end of the scintillator) in blue, and coincidence-mode background in green.}
	\label{1D_bkg}
\end{figure*}

Dark count in SiPMs is the major source of total instrumental background, limiting the instrument sensitivity to detect low energy X-rays. These dark counts are mostly thermally generated carriers which are amplified and mimic the genuine scintillation signal leading to false trigger. Coincidence mode readout from two ends of scintillator would reduce the resultant background significantly. The true scintillation events are more likely to be detected at both ends in coincidence while the random dark counts are largely uncorrelated between two SiPMs. In order to evaluate the impact of 2-sided SiPM readout on the overall background rate, we recorded the events at both ends of the scintillator without illuminating it with the radiation source. The measurements are carried out using the same threshold settings and exposure as those used during source irradiation. Also, a similar analysis procedure is applied to identify coincidence and non-coincidence background events. Figure~\ref{1D_bkg} shows the background light curve with a 2s time binning. It is evident that the background is significantly higher in the non-coincidence mode and SiPM2 consistently exhibits a higher count rate than SiPM1. In coincidence mode, uncorrelated noise events get suppressed leading to the background reduction by a factor of 10 (see Fig.~\ref{1D_bkg}).
The reduced SiPM background also helps reducing the polarimetric background, which can be further suppressed by implementing triple coincidence between the scatterer and both ends of the absorbers. This approach can significantly improve the polarimetric sensitivity of the instrument compared to the previous version of CXPOL.

\section{Summary and Future Work}\label{sec:summary}
In this manuscript, we presented the design and characterisation of the initial prototype of a one-dimensional position sensitive scintillator detector which can be used as absorber detectors for a Compton hard X-ray spectro-polarimeter. The detector consists of a 100$\times$20$\times$5 $\mathrm{mm^3}$ NaI (Tl) scintillator readout by SiPMs on two 20$\times$5 $\mathrm{mm^2}$ ends, providing the energy and one-dimensional position measurement of the incident X-rays. The performance of the detector is assessed with 59.54 keV X-rays from Am$^{241}$ source irradiated along the length of the scintillator. It is observed that peak channel of the line is maximum at the ends of the crystal and reduces towards the centre, a trend consistent with the expected light output at the two ends for interactions along the detector. Enforcing coincidence between the two ends reduces the background significantly and the relative detection efficiency in coincident mode is the highest at the centre of the detector. The results show that the detector has reasonably high detection efficiencies ($>$ 60$\%$) at 60 keV along the length of the detector with coincident
two-end readout. The achieved mean energy resolution is 35$\%$ at 59.54 keV and average spatial resolution along the detector length is $\sim$ 1.5 cm. Position and energy measurements in this detector when used as absorber detectors of a Compton polarimeter, better estimates on the incident photon energy can be obtained allowing spectro-polarimetric studies.

Characterisation of this initial prototype detector also brings up a few aspects of possible further improvement in its design and readout system. The current detector system is shown to be efficient in detecting X-rays of 60 keV along its length in coincident mode; however, this needs to be extended down to $\sim$ 20 keV to make them best suited for Compton polarimeters. 
One aspect for improvement is the optical coupling of the scintillator with the SiPMs, with a better design of the enclosure. Another possible improvement is the use of relatively faster readout electronics, which would result in less integration of SiPM dark counts. This shall be explored with the use of Application Specific
Integrated Circuits (ASICs) meant for SiPM readout as they will also allow readout of multiple detectors simultaneously, which would be needed in a polarimetric configuration. Such an upgrade would significantly improve the timing resolution, allowing for more effective background rejection through tighter coincidence windows. Furthermore, optimizing the front-end electronics is expected to lower the trigger threshold, potentially extending the polarimetric sensitivity down to $\sim$ 20 keV. It is also worth exploring faster scintillators than NaI (Tl), such as GAGG or CeBr${_3}$, which have comparable or higher light yield with much faster scintillation decay constants. Another aspect for optimisation is the bias voltage and operating temperatures of the SiPMs, which will have impact on parameters
such as the gain and dark current of the SiPM, and thus on the X-ray detection sensitivity. It is planned to procure hermetically sealed detectors with quartz windows separately, allowing the SiPMs to be assembled in-house. This will enable SiPMs calibration using known sources prior to final assembly, mitigating any uncertainties in the estimation of the relative gains of the sensors. To ensure long-term stability and reliability of the measurements, it is also planned to validate the detector module’s hermeticity for space-borne use through rigorous thermal cycling and high-humidity environmental tests. These aspects are planned to incorporate in the next version of the detectors being developed for a CXPOL type focal plane Compton polarimeter as well as a collimated Compton X-ray polarimeter instrument unit, which is being considered for a demonstration fight with a small satellite platform.

\begin{acknowledgements}

This research is supported by the Physical Research Laboratory, Ahmedabad, Department of Space, Government of India.	
\end{acknowledgements}

\section*{Conflict of interest}
The authors declare that they have no conflict of interest.

\section*{Availability of data and material}
The instrument characterization data presented in this paper may be made available on reasonable request.
\section*{Code availability}
Not applicable.

\bibliographystyle{spbasic}      


\end{document}